\documentclass[11pt]{article}

\usepackage{amsmath,amssymb}
\usepackage{slashed}
\usepackage{xcolor} 
\usepackage{graphicx}
\usepackage{url}
\usepackage{cite}

\begin{document}

\vspace*{-2cm}
\begin{flushright}
\vspace*{2mm}
\today
\end{flushright}

\begin{center}
\vspace*{15mm}

\vspace{1cm}
{\Large \bf 
Vector-like leptons: Higgs decays and collider phenomenology
} \\
\vspace{1cm}

{\bf A. Falkowski$^{a}$, D. M. Straub$^{b}$ and A. Vicente$^{c}$}

 \vspace*{.5cm} 
$^{a}$ Laboratoire de Physique Th\'eorique, CNRS -- UMR 8627, \\
Universit\'e de Paris-Sud 11, F-91405 Orsay Cedex, France
\vspace*{.2cm} 

$^{b}$ Excellence Cluster Universe, Technische Universit\"at M\"unchen,\\
Boltzmannstr.~2, 85748~Garching, Germany
\vspace*{.2cm} 

$^{c}$ IFPA, Dep. AGO, Universit\'e de Li\`ege, \\
Bat B5, Sart-Tilman B-4000 Li\`ege 1, Belgium

\end{center}

\vspace*{10mm}
\begin{abstract}\noindent\normalsize
We study the impact of heavy vector-like leptons on several
observables in collider and low-energy physics. These states, present
in many well-motivated extensions of the Standard Model, can induce
lepton flavour violation and non-standard Higgs decays.  We study
these effects in an effective model inspired by the composite Higgs
scenario.  After deriving bounds on the mass and production
cross-section of the vector-like states using recent LHC data on
multilepton searches, we discuss the modification of the Higgs decays
to dilepton, diphoton and $Z\gamma$ final states as well as low-energy
observables like radiative lepton decays, the anomalous magnetic
moment of the muon and the electric dipole moment of the electron.  We
find several interesting correlations.  In particular, we show that
branching fractions of lepton flavour-violating Higgs decays at an
observable level are prohibited due to the strong bounds on the
radiative lepton decays.

\end{abstract}

\vspace*{3mm}

\section{Introduction} \label{sec:intro}

Since the start up of the Large Hadron Collider (LHC) in 2008, the
performance of the collider and the results already achieved have been
outstanding. The data collected by the ATLAS and CMS collaborations
have provided conclusive evidence for the existence of the Higgs boson
and put very stringent constraints in many new physics (NP)
scenarios. Well-motivated frameworks, like supersymmetry or extra
dimensions, have been searched with dedicated efforts, but so far no
sign of them has been found.

The discovery of a Higgs boson \cite{Aad:2012tfa,Chatrchyan:2012ufa}
with $m_h \approx 126$~GeV not only gave us the final missing piece in
the Standard Model (SM), but also opened a window to search for new
physics. So far all data seem to be consistent with the 126 GeV
particle being the SM Higgs boson (see
e.g. \cite{Falkowski:2013dza,Giardino:2013bma,Ellis:2013lra,Djouadi:2013qya}),
but the possibility of deviations from the expected SM properties
still remains and this has encouraged many studies in this direction.
For example, new fermions with masses near the weak scale may alter
the Higgs properties.  Although a chiral 4th generation has already
been excluded by the LHC (precisely due to the non-observation of an
enhancement in the Higgs production cross-section), the existence of
one or more vector-like families is a perfectly valid possibility. The
contribution of these new states to Higgs radiative decays can lead to
observable effects at the LHC.

In this paper we concentrate on vector-like leptons.  
 More precisely, we are interested in models in
which the SM charged leptons get their masses due to their mixing with
massive vector-like states. This scenario allows for a simple
understanding of the mass hierarchy among different generations and,
in fact, it is common to many well-motivated extensions of the SM,
such as composite Higgs models \cite{KerenZur:2012fr,Redi:2013pga} and
warped extra dimensions \cite{Contino:2006nn,Agashe:2008fe}.

We explore the phenomenology of vector-like leptons on several
fronts. We will do so by means of a generic setup that covers a wide
class of models in the literature. First of all, we need to determine
the allowed parameter space when LHC data is taken into account. Once
pair-produced, mostly in Drell-Yan processes \cite{Frampton:1999xi},
vector-like leptons will decay to final states including gauge bosons
and charged leptons or neutrinos. Although there are no dedicated
searches for these states at the LHC, generic searches for an
anomalous production of multilepton final states
\cite{CMS-PAS-SUS-13-002} can be recast in order to set bounds on
their mass and production cross-sections.

As mentioned above, one of the main goals of our paper is to determine
the impact of the vector-like leptons on Higgs decays.
Here, we put a special focus
focus on the lepton flavour violating (LFV) Higgs decays $h \to e_i
e_j$, that are a good example of LFV at colliders, a possibility less
explored than LFV at low-energy experiments. Although several authors have
been attracted by these non-standard Higgs decays
\cite{Blankenburg:2012ex,Harnik:2012pb,Davidson:2012ds}, a fully-fledged analysis within a
model with vector-like leptons is still lacking.
We also discuss the possible effects of vector-like leptons on the  the $h\to\gamma\gamma$ and $Z \gamma$ decay rates. 
Especially the $h\to\gamma\gamma$ decay has received considerable attention
in the recent literature  (see e.g.\ \cite{ArkaniHamed:2012kq,Kearney:2012zi,Almeida:2012bq,Carmona:2013cq,Altmannshofer:2013zba,Ishiwata:2011hr})
due to the possibility of a significant enhancement of this mode.
After the diphoton rate has evolved into a value compatible with the SM, the
excitement about $h \to \gamma \gamma$ has clearly
decreased. Nevertheless, it still remains as one of the places where
NP might show up.

Although neutrino flavour oscillations constitute by themselves the
proof that lepton flavour is not conserved, flavour violation in
processes involving charged leptons has never been observed. This is
well understood in the SM (minimally extended to include neutrino
masses), where one expects tiny LFV rates. However, the situation
changes dramatically when vector-like leptons are considered
\cite{Redi:2013pga, Ishiwata:2013gma}. In the absence of flavour
symmetries, the vector-like leptons will induce LFV processes, such as
$e_i \to e_j \gamma$, with potentially large rates. Similarly,
low-energy observables such as $g_\mu-2$ or the electron EDM are also
affected by the presence of vector-like leptons \cite{Kannike:2011ng,Dermisek:2013gta}. We will show that
many of these observables are clearly correlated. This will allow us
to make definite predictions, which in turn will put this NP scenario
under experimental test.

The paper is organized as follows: in sec.~\ref{sec:model} we describe
our setup. After a general review of the experimental situation, we
study the collider phenomenology of vector-like leptons in
sec.~\ref{sec:exp}. In sec.~\ref{sec:pheno}, we present our results,
with special emphasis on correlations among observables, and give the
main predictions of this paper. Finally, we summarize our results and
conclude in sec.~\ref{sec:conclusions}.

\section{The model} \label{sec:model}

In this paper we are interested in the class of models where the masses of SM charged leptons are generated via mixing with heavy vector-like leptons. 
We introduce 3 generations of chiral leptons,  $l^i_L=(\nu_L^i,e_L^i)$, $e_R^i$, $i=1\dots 3$,  and 3 generations of vector-like leptons $L^i = (N^i,E^i)$, $\tilde E^i$, transforming as $2_{-1/2}$ and $1_{-1}$ under the electroweak gauge group.  
The vector-like leptons acquire masses from  Dirac mass terms and from Yukawa couplings with the Higgs boson: 
\begin{equation}
{\cal L}_{F,c} = -   M \left ( \bar L C_L  L +   \bar E C_R E   \right) 
- \left ( \bar L_L Y \tilde E_R H +  \bar L_R \tilde Y \tilde E_L  H + {\rm h.c.} \right ). 
\end{equation}
Here $C_L$, $C_R$, $Y$, $\tilde Y$ are $3 \times 3$ matrices in the
generation space, and we find it convenient to isolate a common scale
$M$ from the vector-like mass terms.  We can always rotate the basis
such that $C_{L,R}$ are diagonal, however $Y$ and $\tilde Y$ are in
general flavour non-diagonal and can induce flavour violating
transitions.  If $M \gg v$ and the elements of $C_{L,R}$ are ${\cal
  O}(1)$ then $M$ sets the mass scale of the heavy leptons.  The
chiral leptons do not acquire mass directly, but only via the mixing
with the vector-like leptons:
\begin{equation}
{\cal L}_\text{mix} =  M \left ( \bar l_L   \lambda_l L_R  +   \overline{\tilde  E_L} \lambda_e   e_R  \right ) +  {\rm h.c.} 
\end{equation}
Here $\lambda_i$ are $3\times 3$ matrices in the generation space. 

This set-up is inspired by composite Higgs models. In this framework, the chiral fermions correspond to the elementary states that mix with fermionic resonances in the composite sector. The Higgs field belongs to the composite sector and therefore has no direct couplings with the elementary fermions.

\begin{figure}[tb]
\centering
\includegraphics[width=0.37\linewidth]{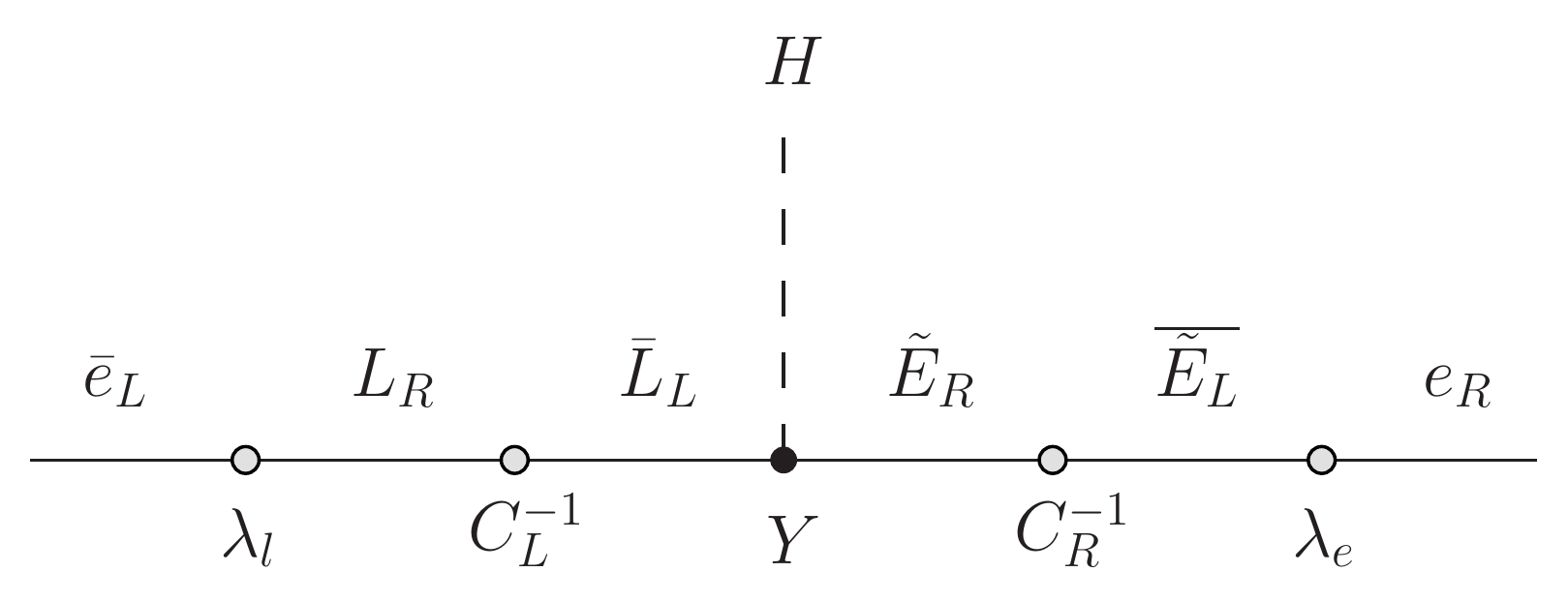}
\includegraphics[width=0.62\linewidth]{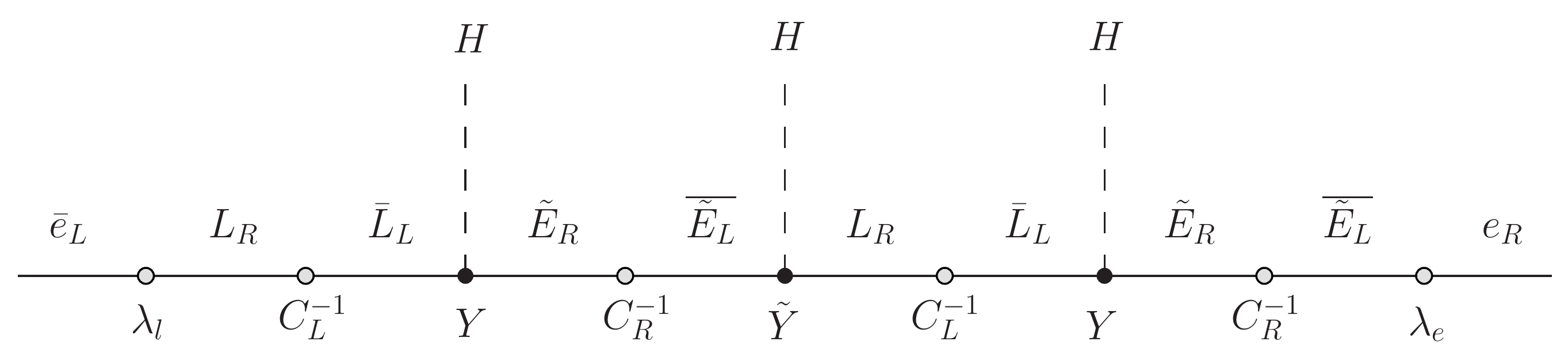}
\caption{Diagramatic illustration of the effective Yukawa coupling.}
\label{fig:mass_terms}
\end{figure}
To understand the structure of the low-energy mass matrix of the charged leptons it is instructive to integrate out the vector-like fields $E$ and $\tilde E$. 
At the zero derivative level one obtains
\begin{equation}
{\cal L}_\text{eff} =  
- \frac{v}{\sqrt 2} \bar e_L   Y_{\rm eff } e_R    +  {\rm h.c.}  
\qquad
Y_{\rm eff} =  \lambda_l C_L^{-1} \left[ Y  +  \frac{v^2 }{ 2 M^2}  Y C_R^{-1} \tilde Y C_L^{-1} Y   + \dots   \right ] C_R^{-1}\lambda_e  \, .
\label{eq:Yeff}
\end{equation}
This can be illustrated diagramatically, see Fig. \ref{fig:mass_terms}. 
Clearly, the effective Yukawa couplings of the chiral leptons are proportional to the mixing parameters $\lambda_{l,e}$. 
By choosing hierarchical $\lambda$'s  it is possible to generate  hierarchies of lepton masses and mixings, even if the Yukawa couplings are anarchic. 
Note that at the leading order the SM lepton masses depend on  the Yukawa matrix $Y$ while the contribution of $\tilde Y$  is suppressed by $v^2/M^2$. 

Similarly, working at the first derivative level one finds the Higgs couplings to leptons. These can be written as
\begin{equation}
{\cal L}_\text{eff} =  
- \frac{h }{ \sqrt 2} \bar e_L   c_{\rm eff } e_R    +  {\rm h.c.}  
\qquad
c_{\rm eff} =  Y_{\rm eff} + \frac{v^2}{M^2} \lambda_l C_L^{-1} Y C_R^{-1} \tilde Y C_L^{-1} Y C_R^{-1}\lambda_e  \, ,
\label{eq:ceff}
\end{equation}
where we have neglected terms of order $\mathcal{O}(\lambda^4)$. The first term contains flavour diagonal Higgs couplings, similar to the ones in the SM. The second term is in general non-diagonal. Note that it vanishes when $\tilde Y = 0$. Moreover, the flavour violating piece also contains a $v^2/M^2$ suppression.

We have not discussed neutrino masses so far. These can be accounted for by adding right-handed neutrinos (and their corresponding vector-like analogs) to our simple setup. Dirac and Majorana neutrino masses and be easily introduced in this way, see \cite{Agashe:2008fe,KerenZur:2012fr,Redi:2013pga,Iyer:2012db}. Since extensions in this direction do not have any impact on charged lepton flavour violation, we choose not to discuss this issue any further.

\section{Current experimental situation and future projects} \label{sec:exp}

In this section we review the current bounds and prospects of experimental searches for LFV and vector-like leptons.

\subsection{Bounds on lepton flavour violation} \label{subsec:LFVexp}

The signatures of LFV processes are being searched for in multiple
experiments.  In table~\ref{tab:sensi} we collect the present bounds
and expected near-future sensitivity to the branching fraction of LFV
lepton decays that may play a role in constraining models with
vector-like leptons.  Typically, the most stringent constraints on
these models come from the limits on the $\mu \to e \gamma$ radiative
decay.  Recently, the MEG collaboration published a new limit,
$\text{Br}(\mu \to e \gamma) < 5.7 \cdot 10^{-13}$, obtained from an
analysis of the 2009-2011 data \cite{Adam:2013mnn}.  Future upgrades
may reach the sensitivity of about $6 \cdot 10^{-14}$ after 3 years of
acquisition time \cite{Baldini:2013ke}.  Limits on LFV radiative
$\tau$ decays are less stringent, although they may be relevant in
scenarios where LFV is Higgs mediated.  The future generation of $B$
factories, in particular Belle II, will be able to shed new light on
$\tau$ LFV decays \cite{O'Leary:2010af,Hayasaka:2013dsa}.  The 3-body
decays of charged leptons, $e_i \to 3 e_j$, and the neutrinoless
conversion in muonic atoms, can also be relevant in certain classes of
models.  However, these decays are suppressed compared to the
radiative decays in models where the LFV amplitudes are dominated by
dipole-type diagrams, as is the case in the model we study here.

\begin{table}[tb!]
\centering
\begin{tabular}{|c|c|c|}
\hline
LFV Process & Present Bound & Future Sensitivity  \\
\hline
$\mu \to e \gamma$ & $5.7 \times 10^{-13}$ \cite{Adam:2013mnn} & $6 \cdot 10^{-14}$ \cite{Baldini:2013ke}  \\
$\tau \to e \gamma$ & $3.3 \times 10^{-8}$ \cite{Aubert:2009ag}& $\sim 10^{-8}-10^{-9}$ \cite{Hayasaka:2013dsa}\\
$\tau \to \mu \gamma$ & $4.4 \times 10^{-8}$ \cite{Aubert:2009ag}& $\sim 10^{-8}-10^{-9}$ \cite{Hayasaka:2013dsa} \\
$\mu \to 3 e$ & $1.0 \times 10^{-12}$\cite{Bellgardt:1987du} & $\sim 10^{-16}$ \cite{Blondel:2013ia}\\
$\tau \to 3 e$ & $2.7\times10^{-8}$\cite{Hayasaka:2010np} & $\sim 10^{-9}-10^{-10}$ \cite{Hayasaka:2013dsa}  \\
$\tau \to 3 \mu$ & $2.1\times10^{-8}$\cite{Hayasaka:2010np} & $\sim 10^{-9}-10^{-10}$ \cite{Hayasaka:2013dsa}  \\
$\mu^-$, Au $\to$ $e^-$, Au & $7.0 \times 10^{-13}$ \cite{Bertl:2006up} & $ $  \\
$\mu^-$, Ti $\to$ $e^-$, Ti & $4.3 \times 10^{-12}$ \cite{Dohmen:1993mp} & $\sim 10^{-18}$ \cite{PRIME} \\
\hline
\end{tabular}
\caption{Current experimental bounds and future sensitivities for some
  low-energy LFV observables.}
\label{tab:sensi}
\end{table}

The recent discovery of the Higgs boson opens the possibility of
searching for LFV Higgs decays $h \to e_i e_j$, with $i \ne j$.
A significant branching fraction for these decays can be compatible
with constraints from other LFV processes
\cite{Blankenburg:2012ex,Harnik:2012pb,Davidson:2012ds}.  Although no
official limits on the branching fractions exist at this moment, a
recast of the $h \to \tau \tau$ search results allows one to place
limits on ${\rm Br} (h \to \tau e/\mu)$ at the level of 10\%
\cite{Harnik:2012pb}.  As we shall see, the branching fractions
predicted in our model are many orders of magnitude smaller, thus
these limits currently play no role in constraining our model.

Finally, it is worthwhile to mention in this context the measurements
of anomalous electric and magnetic dipole moments.  Although these
observables are lepton-flavour conserving, the relevant diagrams are
typically generated along with the LFV ones.  There is the
long-standing $\sim 3.5\sigma$ discrepancy between the predicted and
measured anomalous muon magnetic moment, $\delta a_\mu \approx 2.9
\times 10^{-9}$ \cite{Davier:2010nc}, which can be addressed in models
with vector-like leptons.  Currently, the experimental error on
$a_\mu$ is at the level of $0.6 \times 10^{-9}$.  The {\em Muon g-2}
experiment at Fermilab, which should start taking data in 2016, is
expected to reduce that error by a factor of 3.  This will hopefully
clarify whether the anomaly is due to new physics, although reducing
the error on the theoretical prediction, currently of order $0.5
\times 10^{-9}$, may be more difficult.  Another interesting
observable is the electric dipole moment of the electron $d_e$.  In
this case the SM prediction is extremely small $d_e < 10^{-38} \, e \,
{\rm cm} $ (see e.g. \cite{Pospelov:2005pr}), well below the current
experimental sensitivity.  On the other hand, models with TeV scale
vector-like leptons whose couplings contain new CP violating phases
may easily generate $d_e$ at an observable level.  The most recent
limit from the ACME collaboration obtained using the ThO molecules
reads $d_e < 8.9 \times10^{-29} \, e \, {\rm cm} $ \cite{Baron:2013eja}.
This limit is expected to be improved by an order of magnitude in the
near future.

\subsection{Bounds on vector-like leptons} \label{subsec:VLlepexp}

The rate of LFV processes in our model strongly depends on the masses
of vector-like leptons, and the latter can be directly constrained by
collider searches.  The current bounds on vector-like leptons quoted
by the PDG come from the LEP experiment \cite{Beringer:1900zz}.
Assuming a heavy charged lepton decaying to $W \nu$ with the 100\%
branching fraction, the lower bound on its mass is $100.8$ GeV.
Relaxing that assumption could lead to different limits, but in any
case it would not be much larger than $\sim 100$ GeV
\cite{Beringer:1900zz}.  To the best of our knowledge, no experimental
bounds on vector-like leptons have been obtained so far by the LHC
collaborations.  However, it is possible to recast other existing
searches.  Ref.~\cite{Redi:2013pga} used the ATLAS search for type-III
seesaw mediators \cite{ATLAS:2013hma} to constrain composite leptons,
with the resulting lower limit on the mass in the ballpark of
$300$~GeV.  In the rest of this section, we discuss the constraints
coming from the recent CMS multilepton search
\cite{CMS-PAS-SUS-13-002} that uses the full dataset at
$\sqrt{s}=8$~TeV.

We will focus on the situation in which the lowest lying heavy lepton
state is significantly lighter than the others, and assume it is the
only state produced with a non-negligible rate at the LHC.  The
lightest state decays to SM particles (leptons and weak gauge bosons)
which may lead to multilepton final states when some of the gauge
bosons decay leptonically.  We consider the production of both the
lightest charged lepton $E_1$ and the lightest neutral lepton $N_1$.
For the charged one, the pair production cross section depends not
only on the mass but also on the couplings to the Z boson, which in
turn depend on the mixing angles in the charged lepton sector.  Also
the decay branching fractions depend on the mixing angles.  Two
limiting situations can be distinguished:
\begin{itemize}
\item  {\em Singlet}, where the lightest state consists dominantly of a $SU(2)_L$ singlet $\tilde E$ whose coupling to $Z$ is $g_{E_1}^Z = \frac{g_Y^2 }{  \sqrt{g_L^2 + g_Y^2}}$. 
In this case, the only relevant process is the $E_1^+ E_1^-$ pair production, while  $N_1$ is assumed to be heavier and its production is neglected.  
In the singlet case, $E_1^\pm$  can decay  to $Z l^\pm$ or to $W^\pm \nu$. 
The ratio of these two partial widths is given by $2 \cos^2 \theta_w$, leading to ${\rm Br}(E_1 \to Z l) \approx 0.4$, ${\rm Br}(E_1 \to W \nu) \approx 0.6$. 
\item  {\em Doublet}, where the lightest state consists dominantly of a $SU(2)_L$ doublet  $L$.  
It follows that $N_1$ and $E_1$ are approximately degenerate (we collectively denote them as $L_1$). 
The charged lepton couples to the $Z$ boson with strength $g_{E_1}^Z = -\frac{g_L^2 - g_Y^2 }{ 2  \sqrt{g_L^2 + g_Y^2}}$, while the neutral one  with $g_{N_1}^Z = \frac{\sqrt{g_L^2 + g_Y^2} }{ 2 }$. 
There is also  the charged  current interaction of the $E_1$, $N_1$ pair  with strength $g_{E1,N_1}^W = g_L/\sqrt{2}$.  
Thus, apart from the $E_1$ pair production, also the $N_1 \bar N_1$ and $E_1^\mp N_1 (\bar N_1)$ production processes should be taken into account.
In the doublet case, $E_1$ decays to $Z l$ with 100\% branching fraction, while $N_1$ decays to $W l$.   
\end{itemize}
In this paper we present our results only for these limiting two
cases.  Furthermore, for simplicity we assume that $E_1$ and $N_1$ can
decay to only one of the SM lepton flavours.  To estimate the
sensitivity of the CMS multilepton search we used Madgraph 5
\cite{Alwall:2011uj} to simulate the $L_1 \bar L_1$ production
followed by decay $L_1 \to V l$ where $V=W,Z$ and $l$ is a charged or
neutral SM lepton.  The events were then passed to Pythia
\cite{Sjostrand:2006za} for showering and hadronization and to Delphes
\cite{deFavereau:2013fsa} for the CMS detector simulation.  We
repeated the CMS analysis cuts to determine the efficiency $\times$ acceptance
for each of the multilepton categories defined in
\cite{CMS-PAS-SUS-13-002}.  Then we defined the likelihood function
for the SM background + heavy lepton signal (as a function of the
heavy lepton production cross section) to be a simple product of
Poisson likelihood for each category.  In our procedure we ignored the
theoretical errors on the SM background.

\begin{figure} 
\begin{center}
\includegraphics[width=0.45\textwidth]{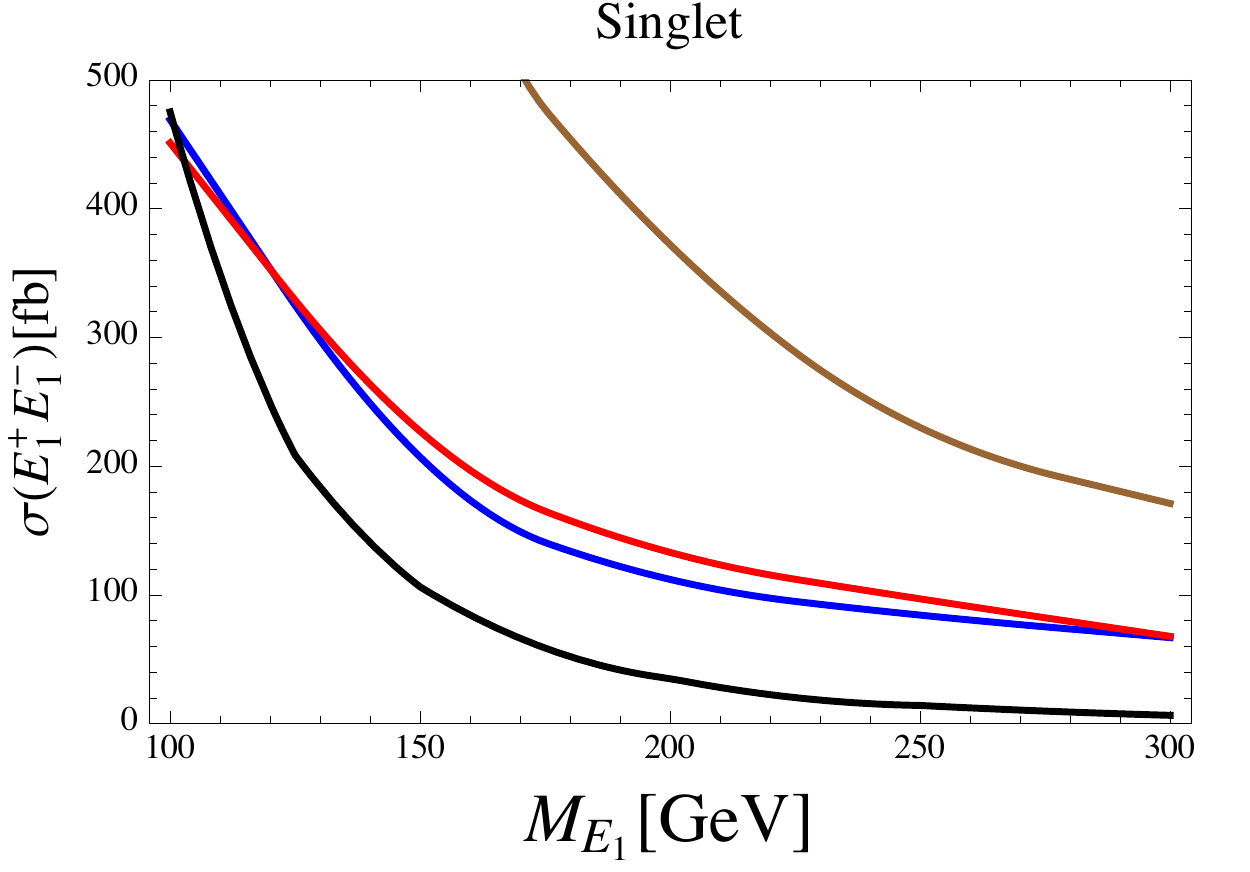}
\quad
\includegraphics[width=0.45\textwidth]{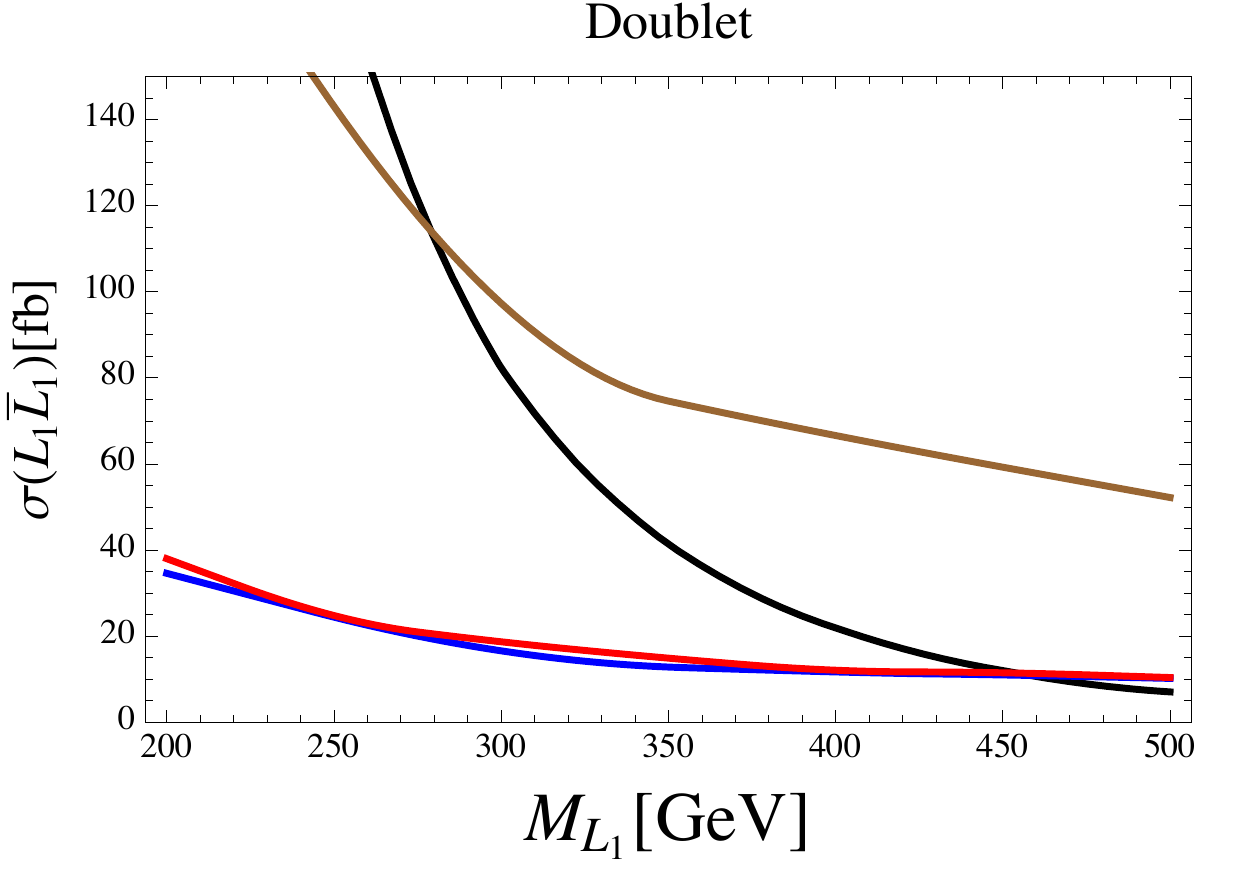}
\end{center}
\caption{Upper limit on the pair production cross section at $\sqrt{s} =8$~TeV LHC of heavy leptons decaying to the electron (red), muon (blue), or tau (brown) SM flavour, for the singlet (left) and doublet (right) cases. The black line shows the production cross section of the heavy lepton via the Drell-Yan processes in our model.
  }
\label{fig:limits}
\end{figure}

Our results are presented in fig.~\ref{fig:limits}.  We give the 95\%
CL limits on the production cross section in the singlet and the
doublet case, separately for heavy leptons decaying to electron, muon,
and tau lepton flavour.  To find the limit on the heavy lepton mass,
this is compared to the production cross section due to the Drell-Yan
processes in our model.  In the singlet case the limits are weak: they
are only slightly above 100 GeV for decays to electrons and muons
(thus marginally improving the LEP limits) and they are below 100 GeV
for decays to tau.  In the doublet case the limits are more stringent,
partly due to the larger production cross section, and partly to the
larger efficiency.  This leads to a lower limit on the mass of $L_1$
around $460$~GeV for the electron and muon decays, and around 280~GeV
for tau decays.

\section{Higgs decays and lepton flavour violation}
\label{sec:pheno}

In this section we first derive approximate analytical expressions for the modification of Higgs decay branching ratios as well as for low-energy observables like charged lepton LFV decays and dipole moments, and finally we perform a numerical analysis.

\subsection{Higgs decays}\label{sec:higgs}

We denote the relative modification of Higgs decay widths compared to the SM as $R(h\to X)\equiv\Gamma(h\to X)/\Gamma(h\to X)_\text{SM}$. In our simple model with vector-like leptons only, this directly corresponds to the experimental signal strengths since corrections to the production cross section are negligible. We note however that in more complete models, e.g.\ composite models, the quark and gauge sectors can generally lead to sizable corrections to the production cross section.

\subsubsection*{Leptonic Higgs decays}

The Higgs decays to electron, muon or $\tau$ pairs are corrected at tree level as
\begin{align}
R(h\to e_i^+e_i^-)
&=
\left|
\frac{c_\text{eff}^{ii}\,v}{\sqrt{2}\,m_{ e_i}}
\right|^2
= \left|1+\frac{\Delta c_\text{eff}^{ii}\,v}{\sqrt{2}\,m_{ e_i}}
\right|^2
\,.
\label{eq:BRhee1}
\end{align}
One can see from eqs.~(\ref{eq:Yeff},~\ref{eq:ceff}) that this expression differs from 1 only for non-zero $\tilde Y^{ij}$.
Interestingly, in this case also lepton flavour violating decays can be generated, which are forbidden in the SM. One can write
\begin{align}
\frac{\Gamma(h\to e_i^\pm e_j^\mp)}{\Gamma(h\to e_j^+ e_j^-)_\text{SM}}
&=
\frac{v^2}{2m_{ e_j}^2}
\left(
|c_\text{eff}^{ij}|^2
+
|c_\text{eff}^{ji}|^2
\right)
\,
\label{eq:BRhee}
\end{align}
for $j>i$.

\subsubsection*{Higgs decays involving photons}

We write the modification of the Higgs to diphoton decay rate as

\begin{align}
R(h\to\gamma\gamma)
&=
\left|
1+\frac{F_H^{\gamma\gamma}}{F_\text{SM}^{\gamma\gamma}}
\right|^2
\,,
\end{align}
where
\begin{align}
F_\text{SM}^{\gamma\gamma}
&=
F_1\!\left(\frac{4 M_W^2}{m_h^2}\right)
+
\frac{4}{3} F_{1/2}\!\left(\frac{4 m_t^2}{m_h^2}\right)
\approx -6.5\,.
\end{align}
Neglecting the mixing between the chiral and vector-like states, $F_H^{\gamma\gamma}$ is dominated by heavy fermion loops and can be written by means of the Higgs low energy theorem as \cite{Carena:2012xa}
\begin{align}
F_H^{\gamma\gamma} \approx \frac{4}{3} v\frac{\partial}{\partial v} \log\left(\det \mathcal M^\dagger \mathcal M\right)
\approx
-\frac{8}{3}\frac{v^2}{M^2}\text{Tr}\left(C_L^{-1} Y C_R^{-1} \tilde{Y} \right)
\label{eq:FH}
\end{align}
where
\begin{equation}
\mathcal M =
\begin{pmatrix}
M C_L & \frac{v}{\sqrt{2}} Y \\
\frac{v}{\sqrt{2}} \tilde Y & M C_R
\end{pmatrix}
\end{equation}
is the heavy block of the charge $-1$ fermion mass matrix. It can be seen from eq.~(\ref{eq:FH}) that $F_H$ can have either sign, so $h\to\gamma\gamma$ can be enhanced or suppressed.

For the $h\to Z\gamma$ decay, 
we focus for simplicity on the case of a single generation that has been worked out e.g.\ in \cite{Altmannshofer:2013zba}. One finds
\begin{align}
R(h\to Z\gamma)
&=
\left|
1+\frac{F_H^{Z\gamma}}{F_\text{SM}^{Z\gamma}}
\right|^2
\,,
\end{align}
where $F_\text{SM}^{Z\gamma}\approx4.8$ and, neglecting the mixing between the chiral and vector-like fermions,
\begin{align}
F_H^{Z\gamma} \approx (1-4 \sin^2 \theta_w)\frac{1}{3}\frac{v^2}{M^2}\, Y \tilde Y + O\!\left(\frac{(C_L-C_E)^2}{(C_L+C_E)^2}\right).
\label{eq:FHZ}
\end{align}
Compared to $h\to \gamma\gamma$, this contribution is accidentally suppressed by $1-4\sin^2 \theta_w\approx 0.08$. In the three-generation case, the correlation between $h\to \gamma\gamma$ and $h\to Z\gamma$ depends additionally on $O(1)$ factors that depend on the flavour structure of the $Y$ and $\tilde Y$, so we do not expect a clear-cut correlation.

\subsection{Low-energy observables}\label{sec:lfv}
\subsubsection*{Radiative lepton decay}

Normalizing the branching ratio of the radiative LFV decay to the semileptonic decay, one has
\begin{align}
\frac{\text{BR}(e_i\to e_j\gamma)}{\text{BR}(e_i\to e_j\nu_i\bar\nu_j)}
&=
\frac{3\alpha}{4\pi G_F^2}
\left(|A_L^{ij}|^2+|A_R^{ij}|^2\right),
\end{align}
where
\begin{align}
A_L^{ij} &=
\frac{v^2}{2 M^2}
\frac{1}{\sqrt{2} \,m_{e_i} v}
\,
\left(\lambda_l C_L^{-1} Y C_R^{-1} \tilde Y C_L^{-1} Y C_R^{-1}\lambda_e\right)_{ij}
\,,
\\
A_R^{ij} &=
\frac{v^2}{2 M^2}
\frac{1}{\sqrt{2} \,m_{e_i} v}
\,
\left(\lambda_l C_L^{-1} Y C_R^{-1} \tilde Y C_L^{-1} Y C_R^{-1}\lambda_e\right)_{ji}
\,.
\end{align}
By comparison to (\ref{eq:ceff}), one sees that these expressions are proportional to the effective flavour-violating Higgs couplings and one can write
\begin{align}
A_L^{ij} &=
\frac{c_\text{eff}^{ij}}{2 \sqrt{2} \,m_{e_i} v}
\,,
&
A_R^{ij} &=
\frac{c_\text{eff}^{ji}}{2 \sqrt{2} \,m_{e_i} v}
\,.
\label{eq:ALAR}
\end{align}
Eqs.~(\ref{eq:BRhee}), (\ref{eq:ALAR}) leads to a perfect correlation between the LFV Higgs decays $h\to e_i e_j$ and the radiative $e_i\to e_j\gamma$ decays,
\begin{equation}
\frac{\text{BR}(h\to e_i e_j)}{\text{BR}(e_i\to e_j\gamma)}
=
\frac{\text{BR}(h\to e_i e_i)_\text{SM}}{\text{BR}(e_i\to e_j\nu_i\bar\nu_j)}
\, \frac{4\pi}{3\alpha}
\end{equation}
Inserting numbers, one obtains
\begin{align}
\text{BR}(h\to \tau \mu) &< 8.6 \times 10^{-6} \, \left[\frac{\text{BR}(\tau\to \mu \gamma)}{4.4 \times 10^{-8}}\right],
\label{eq:corrhtm}
\\
\text{BR}(h\to \tau e) &< 6.2 \times 10^{-6} \, \left[\frac{\text{BR}(\tau\to e \gamma)}{3.3 \times 10^{-8}}\right],
\label{eq:corrhte}
\\
\text{BR}(h\to \mu e) &< 6.7 \times 10^{-14} \, \left[\frac{\text{BR}(\mu\to e \gamma)}{5.7 \times 10^{-13}}\right],
\end{align}
where the branching ratios in square brackets are normalized to the current upper bounds. We conclude that LFV Higgs decay branching ratios are at least four orders of magnitude smaller than the Higgs decays to tau pairs and are thus most probably unobservable.

\subsubsection*{Muon anomalous magnetic moment}

In the flavour conserving case, the dipole operators contribute to the anomalous magnetic moment of the muon, $a_\mu=(g_\mu-2)/2$ \cite{Kannike:2011ng}. One can again write it in terms of $c_\text{eff}$ as
\begin{align}
\delta a_\mu
&\approx
-\frac{m_\mu}{8\sqrt{2}\pi^2 v} 
\,
\text{Re}\,\Delta c_\text{eff}^{\mu\mu}
\,.
\end{align}
Together with eq.~(\ref{eq:BRhee1}), one can then write for small $\Delta c_\text{eff}^{\mu\mu}$
\begin{equation}
R(h\to \mu^+\mu^-) \approx  1-  \left[\frac{\delta a_\mu}{1.6\times 10^{-9}}\right]\,.
\label{eq:corrRmm}
\end{equation}
If the current discrepancy is due to NP, in this model one would thus expect a significant suppression of the $h\to\mu\mu$ branching ratio.

\subsubsection*{Electron electric dipole moment}

The electron EDM is given by
\begin{equation}
\frac{d_e}{e}
\approx
-\frac{1}{16\sqrt{2}\pi^2 v}
\,
\text{Im}\,\Delta c_\text{eff}^{ee}
\,.
\end{equation}
Assuming the flavour structure of the $Y$, $\tilde Y$ and $C_{L,R}$ do not lead to any additional suppression, one expects roughly
\begin{equation}
|\Delta c_\text{eff}^{ee}| \sim \frac{m_e}{m_\mu} |\Delta c_\text{eff}^{\mu\mu}|
\end{equation}
and thus obtains a correlation between $d_e$ and $g_\mu-2$, namely
\begin{align}
d_e &\sim \frac{e}{2}\frac{m_e}{m_\mu}\frac{\delta a_\mu}{m_\mu} \sin\varphi_e
\\
&\approx 5 \times 10^{-25} \,e\,\text{cm} \times\left[\frac{\delta a_\mu}{10^{-9}}\right]  \sin\varphi_e \,,
\end{align}
where $\varphi_e$ is the phase of $\Delta c_\text{eff}^{ee}$.
Consequently, an explanation of the muon $g-2$ anomaly would require a phase below $10^{-4}$ to be in agreement with the recent bound from the ACME experiment $d_e \leq 8.7 \times 10^{-29} \,e\,\text{cm}$. 
For an $O(1)$ phase, one would instead require $M\gtrsim 23 \sqrt{|Y\tilde Y|}$~TeV to fulfill the EDM bound and the contribution to $g_\mu-2$ would be negligible.

\medskip
Before moving on to the numerical analysis, we briefly comment on the dependence of the phenomenology on the choice of fermion representations. Although we have focused on the minimal case of one $SU(2)_L$ doublet and one singlet mixing with the SM leptons, also more complicated possibilities like triplets are conceivable, or bi-doublets under a custodial group $SU(2)_L\times SU(2)_R$ as in many composite models.
In that case, a different $O(1)$ factor can appear in front of $\Delta c_\text{eff}$, i.e.\ the modification of the Higgs coupling with respect to the SM. In addition, a different $O(1)$ factor can multiply all the amplitudes induced by the dipole operator.\footnote{These factors have been computed for several different choices of representations in ref.~\cite{Kannike:2011ng}.}
However, the qualitative correlations between the tree level Higgs decays and the dipole-induced observables will still hold.
The most significant change occurs in the loop-induced $h\to\gamma\gamma$  and $h\to Z\gamma$ decays in the presence of large representations with high charges \cite{Altmannshofer:2013zba}.

\subsection{Numerical analysis}

Having discussed the most important indirect probes of our scenario, we proceed with a numerical analysis, taking into account the direct bounds as discussed in section~\ref{subsec:VLlepexp}.
To this end, we conduct a scan over the model parameters in the following ranges,
\begin{align}
(M C_{L,R})_{ii} &\in [100,1500]\,\text{GeV}
\,,&
\frac{(\lambda_{l,e})_{ii}}{(M C_{L,R})_{ii}} &\in [10^{-5},1]
\,,\\
|Y_{ii}|,|\tilde Y_{ii}| &\in [1/3,3]
\,,&
|Y_{ij}|,|\tilde Y_{ij}| &\in [10^{-5},1]
\,,
\end{align}
where we have scanned the masses and diagonal elements of the Yukawa matrices linearly and the mixings $\lambda_{l,e}$ and the off-diagonal elements of the Yukawa matrices logarithmically, since they can span many orders of magnitude. We have also scanned the phases of the $Y$ and $\tilde Y$ between 0 and $2\pi$.

After imposing that the three charged lepton masses be reproduced correctly, which fixes three of the six parameters in $\lambda_{l,e}$, we remove all points where one of the heavy mass eigenstates violates the direct LEP bound of 100~GeV.
We also impose the LEP bounds on modified $Z$ couplings to charged lepton pairs \cite{ALEPH:2005ab} that constrain the relative modification to the per mille level. This constraint is particularly relevant for the $\tau$, that has the largest mixing with the vector-like leptons.
For the points surviving these constraints, we compute the cross section times branching ratio for the pair production and decay to $W$ or $Z$ and leptons and impose the LHC bounds as discussed in section~\ref{subsec:VLlepexp}. Points that are in violation of this bound are shown in light gray in the following plots, while allowed points are shown in blue.

Finally, we compute all the observables discussed in sections~\ref{sec:higgs} and \ref{sec:lfv}. We stress that we do not employ the approximate analytical formulae, which have been presented above only to get an analytical understanding of the correlations. Rather, we diagonalize the full $9\times9$ mass matrix for charge $-1$ fermion mass eigenstates (and correspondingly for the neutral mass eigenstates) and also compute the loop-induced observables taking into account the full dependence of the loop functions without any approximations.

\begin{figure}
\centering
\includegraphics[width=0.9\textwidth]{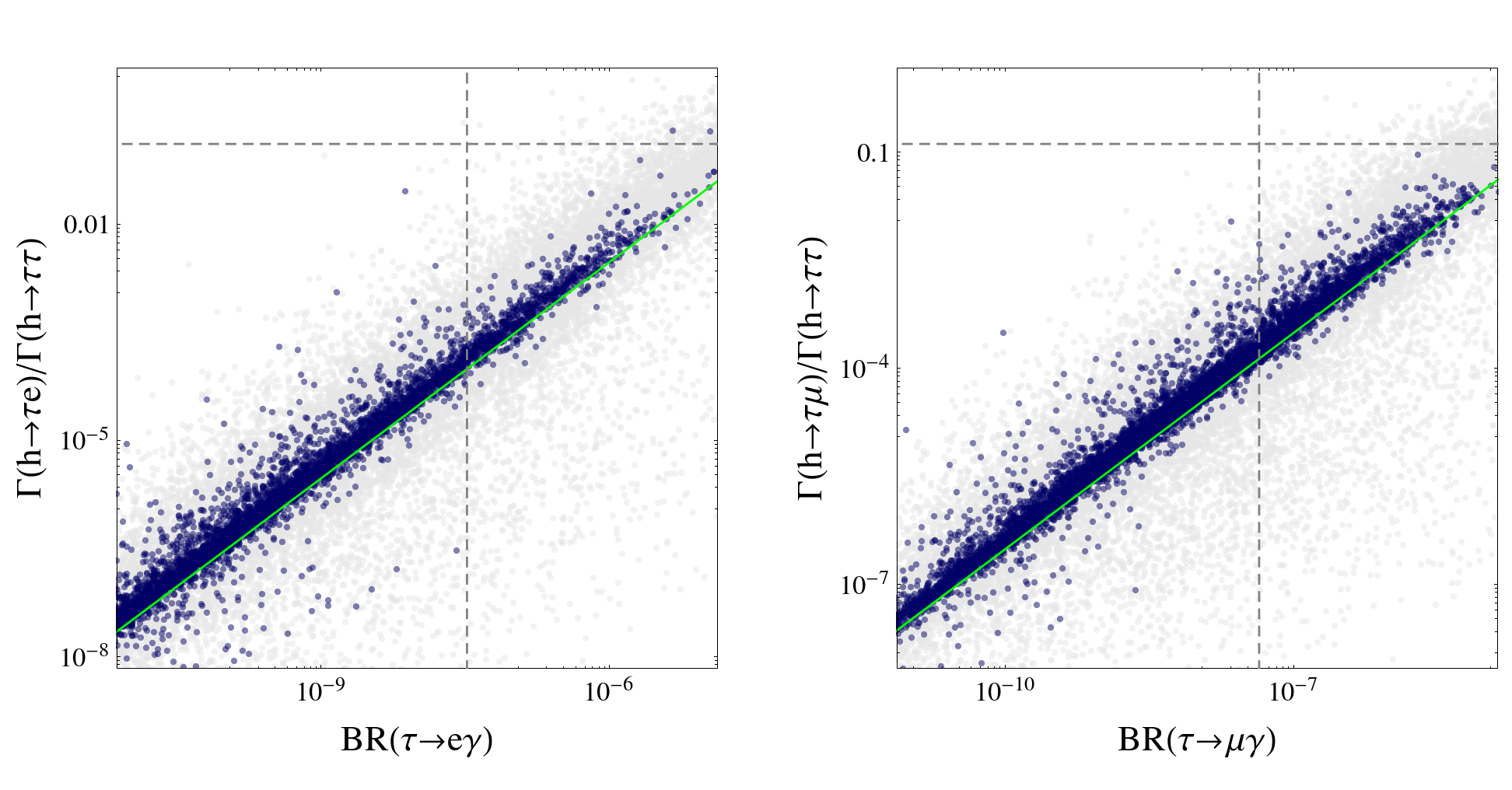}
\caption{Correlation between flavour-violating Higgs decays and radiative lepton decays. Gray points are ruled out by LHC direct searches. The experimentally allowed region is left of and below the dashed lines. The green lines correspond to eqs.~(\ref{eq:corrhte}) and (\ref{eq:corrhtm}).}
\label{fig:scatter1}
\end{figure}

Fig.~\ref{fig:scatter1} shows the results for the correlations between flavour violating Higgs decays and radiative lepton decays. The green lines show the approximate result given in eqs.\ (\ref{eq:corrhtm}) and (\ref{eq:corrhte}) which can be seen to be fulfilled to an excellent precision, in particular once the LHC direct constraints are taken into account. The conclusion that flavour-violating Higgs decays will be unobservable at the LHC in this model is thus robust.

\begin{figure}
\centering
\includegraphics[width=0.9\textwidth]{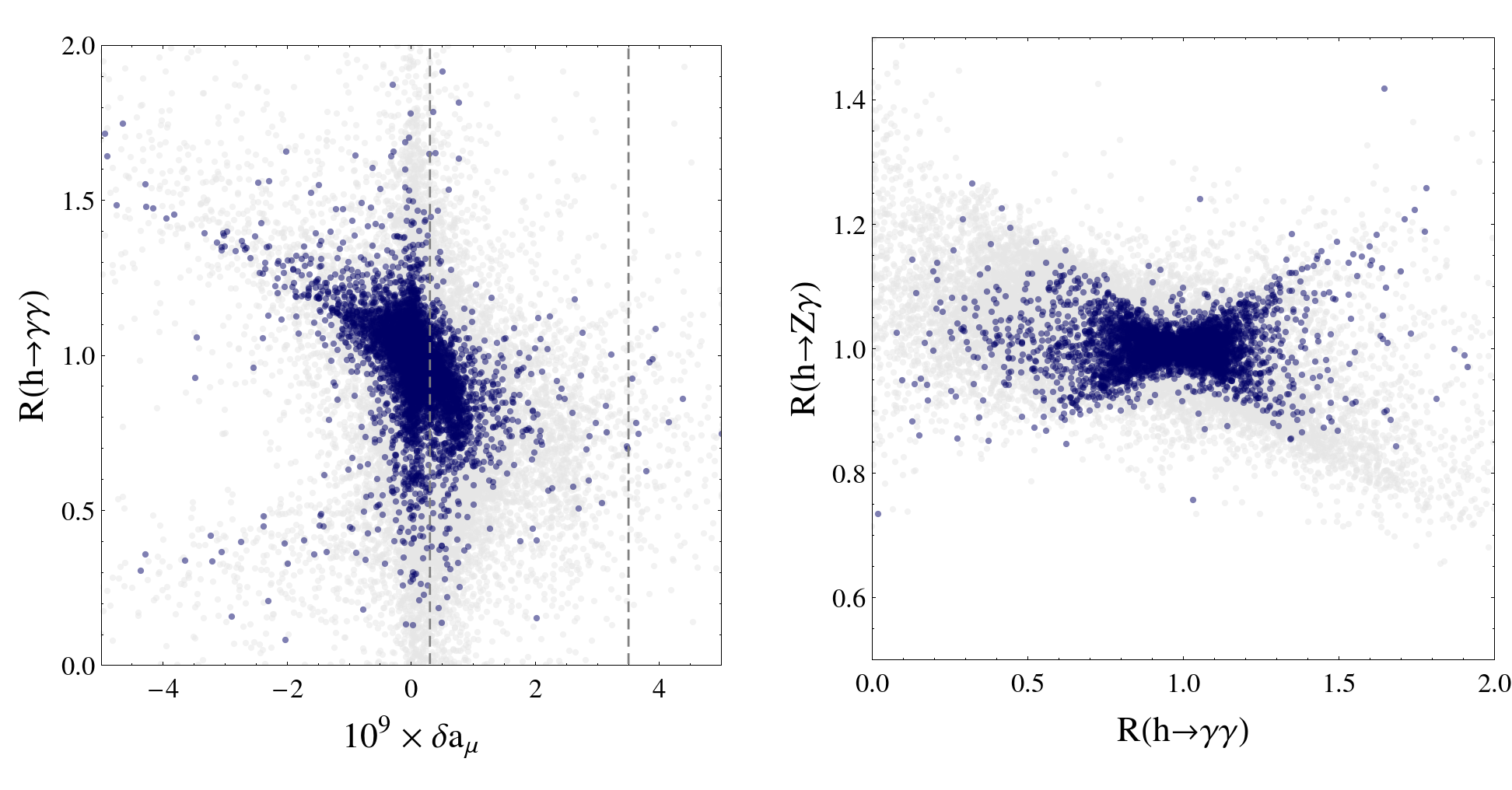}
\caption{Correlation between $h\to\gamma\gamma$, $h\to Z \gamma$ and the muon anomalous magnetic moment. Gray points are ruled out by LHC direct searches. The dashed lines show the $2\sigma$ experimental region.}
\label{fig:scatter2}
\end{figure}

The left-hand panel of fig.~\ref{fig:scatter2} shows the correlation of the Higgs to diphoton decay with the muon anomalous magnetic moment. Although there is now clear correlation in the three-generation flavour-anarchic case we consider, we observe that a solution of the magnetic muon anomaly typically leads to a suppression of $h\to\gamma\gamma$. The right-hand panel shows the correlation between $h\to\gamma\gamma$ and $h\to Z\gamma$. As discussed above, the effects in the latter decay are typically smaller due to the accidental suppression by $(1-4\sin \theta_w^2)$.

\begin{figure}
\centering
\includegraphics[width=0.9\textwidth]{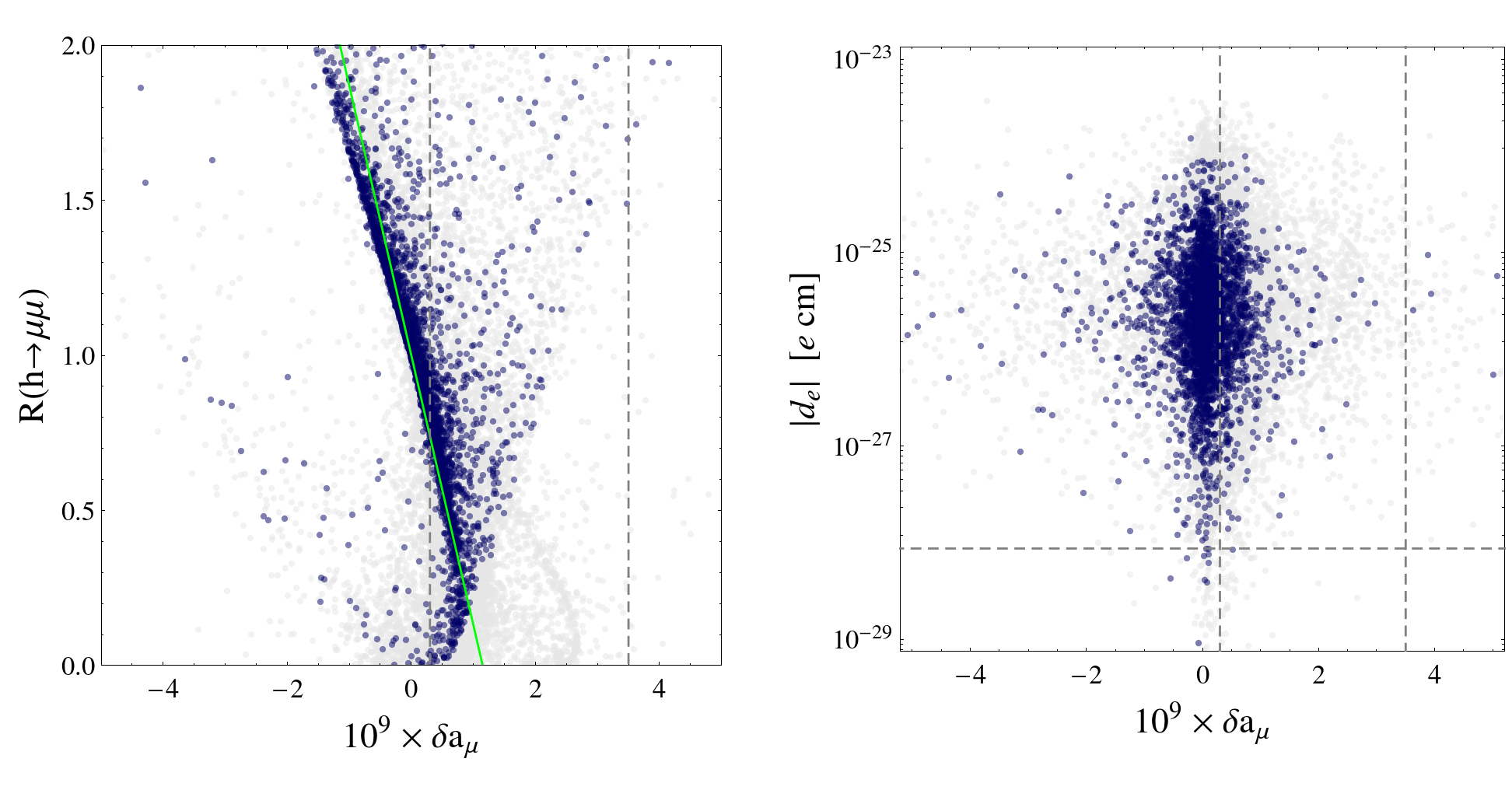}
\caption{Correlation of the muon anomalous magnetic moment with $h\to\mu\mu$ and the electron EDM, respectively.
Gray points are ruled out by LHC direct searches. The dashed lines show the $2\sigma$ experimental region for $\delta a_\mu$ and the 90\% C.L.\ upper bound on $|d_e|$.
The green line corresponds to eq.~(\ref{eq:corrRmm}).
}
\label{fig:scatter3}
\end{figure}

Fig.~\ref{fig:scatter3} shows the correlation between the muon anomalous magnetic moment and the $h\to\mu\mu$ decay as well as the electron EDM.
In the first case, we also show the approximate relation given in eq.~(\ref{eq:corrRmm}) as a solid green line. We can see that this relation is fulfilled to a good precision once the LHC direct constraints are taken into account, but there are also some viable points with sizable positive contribution to $\delta a_\mu$, as preferred by experiment, but no significant modification of the $h\to\mu\mu$ decay rate.
The second plot shows that, as expected from the above discussion, allowing the phases in the Yukawas to be of $O(1)$, one typically produces a too large electron EDM that is ruled out by the bound from the ACME collaboration.

\section{Summary and conclusions} \label{sec:conclusions}

We have explored the phenomenology of vector-like leptons, with
detailed studies of their impact on lepton flavour violation, Higgs
decays and collider physics. These states appear in many well
motivated extensions of the SM, like composite Higgs models, or models
with extra dimensions.  For this reason, we have aimed at a general
effective theory setup so as to obtain conclusion valid in a large
class of these models.

After the discovery of the Higgs boson, the next step is to determine
its properties and look for non-standard signatures. In this paper we
have explored how the presence of vector-like leptons might affect two
Higgs decays channels: (i) $h \to e_i e_j$, with $i \ne j$, and
(ii) $h \to \gamma \gamma$. In the former case, our results clearly
show that the strong upper bounds on $\text{BR}(e_i \to e_j
\gamma)$ preclude any possibility to observe Higgs LFV decays at the
LHC. Therefore, in case these were observed at the LHC, vector-like
leptons would be ruled out as the only source of LFV. On the other
hand, the Higgs diphoton rate can actually be modified due to the new
contributions induced by the vector-like leptons. These do not have a
definite sign and thus $\text{BR}(h \to \gamma \gamma)$ can be either
enhanced or suppressed with respect to the SM prediction. Effects in
$\text{BR}(h \to Z \gamma)$ are typically smaller.

Vector-like leptons also contribute to the dipole operators involved
in $g_\mu-2$. These contributions are found to be correlated with the
corrections to $\text{BR}(h \to \mu \mu)$. The conclusion is that, if
the discrepancy in $g_\mu-2$ is explained thanks to the vector-like
leptons, we expect a significant suppression of the $h \to \mu \mu$
branching ratio.  At the same time, a large enough contribution to
$g_\mu-2$ requires generic CP violating phases in the Yukawa matrices
to be of order $\mathcal{O}(10^{-4})$, otherwise the bound on the
electron EDM would be violated.

In what concerns high-energy signatures of vector-like leptons, we
have focused on their pair production at the LHC.  Using the recent
multilepton search released by the CMS collaboration, we have derived
$95\%$ C.L. limits on the mass of the vector-like lepton as a function
of its production cross-section.  For a charged vector-like lepton
that is a singlet of the SM $SU(2)_L$ group, the CMS search does not
improve on the previous LEP limits of $m_E \gtrsim 100$~GeV.  In contrast,
for a pair of charged and neutral vector-like leptons that form a
$SU(2)_L$ doublet the CMS limits turn out to be non-trivial.  The limits
for a heavy lepton doublet decaying to the $e$ or $\mu$ flavours are
$m_L \gtrsim 450$~GeV.  In the case of decays to the $\tau$ flavour
the limits are less stringent: $m_L \gtrsim 270$~GeV.

\section*{Note added}

While finalizing the manuscript, we became aware of a talk by
M.~Beneke \cite{BenekeTalk} where the connection between LFV Higgs
decays and dipole transitions was discussed in the context of
Randall-Sundrum models.

\section*{Acknowledgements}

We thank Michele Redi and Paride Paradisi for useful discussions.  D.M.S. was supported by
the Advanced Grant EFT4LHC of the European Research Council (ERC), the
Cluster of Excellence {\em Precision Physics, Fundamental Interactions
  and Structure of Matter\/} (PRISMA -- EXC 1098) as well as the
Cluster of Excellence {\em Origin and Structure of the Universe\/} and
thanks the Galileo Galilei Institute for Theoretical Physics for the
hospitality and the INFN for partial support while some of this work
was carried out.  A.V. acknowledges partial support from the ANR
project CPV-LFV-LHC {NT09-508531}.  The work of A.F. was supported by
the ERC advanced grant Higgs@LHC.

\bibliographystyle{JHEP}
\bibliography{CLFV}

\providecommand{\href}[2]{#2}\begingroup\raggedright\begin{thebibliography}{10}

\bibitem{Aad:2012tfa}
{\bf ATLAS Collaboration} Collaboration, G.~Aad et~al., {\it {Observation of a
  new particle in the search for the Standard Model Higgs boson with the ATLAS
  detector at the LHC}},  {\em Phys.Lett.} {\bf B716} (2012) 1--29,
  [\href{http://xxx.lanl.gov/abs/1207.7214}{{\tt arXiv:1207.7214}}].

\bibitem{Chatrchyan:2012ufa}
{\bf CMS} Collaboration, S.~Chatrchyan et~al., {\it {Observation of a new boson
  at a mass of 125 GeV with the CMS experiment at the LHC}},  {\em Phys.Lett.}
  {\bf B716} (2012) 30--61, [\href{http://xxx.lanl.gov/abs/1207.7235}{{\tt
  arXiv:1207.7235}}].

\bibitem{Falkowski:2013dza}
A.~Falkowski, F.~Riva, and A.~Urbano, {\it {Higgs at last}},  {\em JHEP} {\bf
  1311} (2013) 111, [\href{http://xxx.lanl.gov/abs/1303.1812}{{\tt
  arXiv:1303.1812}}].

\bibitem{Giardino:2013bma}
P.~P. Giardino, K.~Kannike, I.~Masina, M.~Raidal, and A.~Strumia, {\it {The
  universal Higgs fit}},  \href{http://xxx.lanl.gov/abs/1303.3570}{{\tt
  arXiv:1303.3570}}.

\bibitem{Ellis:2013lra}
J.~Ellis and T.~You, {\it {Updated Global Analysis of Higgs Couplings}},  {\em
  JHEP} {\bf 1306} (2013) 103, [\href{http://xxx.lanl.gov/abs/1303.3879}{{\tt
  arXiv:1303.3879}}].

\bibitem{Djouadi:2013qya}
A.~Djouadi and G.~Moreau, {\it {The couplings of the Higgs boson and its CP
  properties from fits of the signal strengths and their ratios at the 7+8 TeV
  LHC}},  \href{http://xxx.lanl.gov/abs/1303.6591}{{\tt arXiv:1303.6591}}.

\bibitem{KerenZur:2012fr}
B.~Keren-Zur, P.~Lodone, M.~Nardecchia, D.~Pappadopulo, R.~Rattazzi, et~al.,
  {\it {On Partial Compositeness and the CP asymmetry in charm decays}},  {\em
  Nucl.Phys.} {\bf B867} (2013) 429--447,
  [\href{http://xxx.lanl.gov/abs/1205.5803}{{\tt arXiv:1205.5803}}].

\bibitem{Redi:2013pga}
M.~Redi, {\it {Leptons in Composite MFV}},  {\em JHEP} {\bf 1309} (2013) 060,
  [\href{http://xxx.lanl.gov/abs/1306.1525}{{\tt arXiv:1306.1525}}].

\bibitem{Contino:2006nn}
R.~Contino, T.~Kramer, M.~Son, and R.~Sundrum, {\it {Warped/composite
  phenomenology simplified}},  {\em JHEP} {\bf 0705} (2007) 074,
  [\href{http://xxx.lanl.gov/abs/hep-ph/0612180}{{\tt hep-ph/0612180}}].

\bibitem{Agashe:2008fe}
K.~Agashe, T.~Okui, and R.~Sundrum, {\it {A Common Origin for Neutrino Anarchy
  and Charged Hierarchies}},  {\em Phys.Rev.Lett.} {\bf 102} (2009) 101801,
  [\href{http://xxx.lanl.gov/abs/0810.1277}{{\tt arXiv:0810.1277}}].

\bibitem{Frampton:1999xi}
P.~H. Frampton, P.~Hung, and M.~Sher, {\it {Quarks and leptons beyond the third
  generation}},  {\em Phys.Rept.} {\bf 330} (2000) 263,
  [\href{http://xxx.lanl.gov/abs/hep-ph/9903387}{{\tt hep-ph/9903387}}].

\bibitem{CMS-PAS-SUS-13-002}
{\bf CMS} Collaboration, {\it {A search for anomalous production of events with
  three or more leptons using 19.5/fb of sqrt(s)=8 TeV LHC data}},  Tech. Rep.
  CMS-PAS-SUS-13-002, CERN, Geneva, 2013.

\bibitem{Blankenburg:2012ex}
G.~Blankenburg, J.~Ellis, and G.~Isidori, {\it {Flavour-Changing Decays of a
  125 GeV Higgs-like Particle}},  {\em Phys.Lett.} {\bf B712} (2012) 386--390,
  [\href{http://xxx.lanl.gov/abs/1202.5704}{{\tt arXiv:1202.5704}}].

\bibitem{Harnik:2012pb}
R.~Harnik, J.~Kopp, and J.~Zupan, {\it {Flavor Violating Higgs Decays}},  {\em
  JHEP} {\bf 1303} (2013) 026, [\href{http://xxx.lanl.gov/abs/1209.1397}{{\tt
  arXiv:1209.1397}}].

\bibitem{Davidson:2012ds}
S.~Davidson and P.~Verdier, {\it {LHC sensitivity to the decay $h \to \tau^\pm
  \mu^\mp$}},  {\em Phys.Rev.} {\bf D86} (2012) 111701,
  [\href{http://xxx.lanl.gov/abs/1211.1248}{{\tt arXiv:1211.1248}}].

\bibitem{ArkaniHamed:2012kq}
N.~Arkani-Hamed, K.~Blum, R.~T. D'Agnolo, and J.~Fan, {\it {2:1 for Naturalness
  at the LHC?}},  {\em JHEP} {\bf 1301} (2013) 149,
  [\href{http://xxx.lanl.gov/abs/1207.4482}{{\tt arXiv:1207.4482}}].

\bibitem{Kearney:2012zi}
J.~Kearney, A.~Pierce, and N.~Weiner, {\it {Vectorlike Fermions and Higgs
  Couplings}},  {\em Phys.Rev.} {\bf D86} (2012) 113005,
  [\href{http://xxx.lanl.gov/abs/1207.7062}{{\tt arXiv:1207.7062}}].

\bibitem{Almeida:2012bq}
L.~G. Almeida, E.~Bertuzzo, P.~A. Machado, and R.~Z. Funchal, {\it {Does $H \to
  \gamma \gamma$ Taste like vanilla New Physics?}},  {\em JHEP} {\bf 1211}
  (2012) 085, [\href{http://xxx.lanl.gov/abs/1207.5254}{{\tt
  arXiv:1207.5254}}].

\bibitem{Carmona:2013cq}
A.~Carmona and F.~Goertz, {\it {Custodial Leptons and Higgs Decays}},  {\em
  JHEP} {\bf 1304} (2013) 163, [\href{http://xxx.lanl.gov/abs/1301.5856}{{\tt
  arXiv:1301.5856}}].

\bibitem{Altmannshofer:2013zba}
W.~Altmannshofer, M.~Bauer, and M.~Carena, {\it {Exotic Leptons: Higgs, Flavor
  and Collider Phenomenology}},  \href{http://xxx.lanl.gov/abs/1308.1987}{{\tt
  arXiv:1308.1987}}.

\bibitem{Ishiwata:2011hr}
K.~Ishiwata and M.~B. Wise, {\it {Higgs Properties and Fourth Generation
  Leptons}},  {\em Phys.Rev.} {\bf D84} (2011) 055025,
  [\href{http://xxx.lanl.gov/abs/1107.1490}{{\tt arXiv:1107.1490}}].

\bibitem{Ishiwata:2013gma}
K.~Ishiwata and M.~B. Wise, {\it {Phenomenology of heavy vector-like leptons}},
   \href{http://xxx.lanl.gov/abs/1307.1112}{{\tt arXiv:1307.1112}}.

\bibitem{Kannike:2011ng}
K.~Kannike, M.~Raidal, D.~M. Straub, and A.~Strumia, {\it {Anthropic solution
  to the magnetic muon anomaly: the charged see-saw}},  {\em JHEP} {\bf 1202}
  (2012) 106, [\href{http://xxx.lanl.gov/abs/1111.2551}{{\tt
  arXiv:1111.2551}}].

\bibitem{Dermisek:2013gta}
R.~Dermisek and A.~Raval, {\it {Explanation of the Muon g-2 Anomaly with
  Vectorlike Leptons and its Implications for Higgs Decays}},  {\em Phys.Rev.}
  {\bf D88} (2013) 013017, [\href{http://xxx.lanl.gov/abs/1305.3522}{{\tt
  arXiv:1305.3522}}].

\bibitem{Iyer:2012db}
A.~M. Iyer and S.~K. Vempati, {\it {Lepton Masses and Flavor Violation in
  Randall Sundrum Model}},  {\em Phys.Rev.} {\bf D86} (2012) 056005,
  [\href{http://xxx.lanl.gov/abs/1206.4383}{{\tt arXiv:1206.4383}}].

\bibitem{Adam:2013mnn}
{\bf MEG Collaboration} Collaboration, J.~Adam et~al., {\it {New constraint on
  the existence of the $\mu^+ \to e^+ \gamma$ decay}},  {\em Phys.Rev.Lett.}
  {\bf 110} (2013) 201801, [\href{http://xxx.lanl.gov/abs/1303.0754}{{\tt
  arXiv:1303.0754}}].

\bibitem{Baldini:2013ke}
A.~Baldini, F.~Cei, C.~Cerri, S.~Dussoni, L.~Galli, et~al., {\it {MEG Upgrade
  Proposal}},  \href{http://xxx.lanl.gov/abs/1301.7225}{{\tt arXiv:1301.7225}}.

\bibitem{O'Leary:2010af}
{\bf SuperB Collaboration} Collaboration, B.~O'Leary et~al., {\it {SuperB
  Progress Reports -- Physics}},  \href{http://xxx.lanl.gov/abs/1008.1541}{{\tt
  arXiv:1008.1541}}.

\bibitem{Hayasaka:2013dsa}
{\bf Belle and Belle II Collaborations} Collaboration, K.~Hayasaka, {\it
  {Results and prospects on lepton flavor violation at Belle/Belle II}},  {\em
  J.Phys.Conf.Ser.} {\bf 408} (2013) 012069.

\bibitem{Aubert:2009ag}
{\bf BaBar Collaboration} Collaboration, B.~Aubert et~al., {\it {Searches for
  Lepton Flavor Violation in the Decays $tau^\pm \to e^\pm \gamma$ and $tau^\pm
  \to mu^\pm \gamma$}},  {\em Phys.Rev.Lett.} {\bf 104} (2010) 021802,
  [\href{http://xxx.lanl.gov/abs/0908.2381}{{\tt arXiv:0908.2381}}].

\bibitem{Bellgardt:1987du}
{\bf SINDRUM Collaboration} Collaboration, U.~Bellgardt et~al., {\it {Search
  for the Decay $mu^+ \to e^+ e^+ e^-$}},  {\em Nucl.Phys.} {\bf B299} (1988)
  1.

\bibitem{Blondel:2013ia}
A.~Blondel, A.~Bravar, M.~Pohl, S.~Bachmann, N.~Berger, et~al., {\it {Research
  Proposal for an Experiment to Search for the Decay ${\mu} \to eee$}},
  \href{http://xxx.lanl.gov/abs/1301.6113}{{\tt arXiv:1301.6113}}.

\bibitem{Hayasaka:2010np}
K.~Hayasaka, K.~Inami, Y.~Miyazaki, K.~Arinstein, V.~Aulchenko, et~al., {\it
  {Search for Lepton Flavor Violating Tau Decays into Three Leptons with 719
  Million Produced Tau+Tau- Pairs}},  {\em Phys.Lett.} {\bf B687} (2010)
  139--143, [\href{http://xxx.lanl.gov/abs/1001.3221}{{\tt arXiv:1001.3221}}].

\bibitem{Bertl:2006up}
{\bf SINDRUM II Collaboration} Collaboration, W.~H. Bertl et~al., {\it {A
  Search for muon to electron conversion in muonic gold}},  {\em Eur.Phys.J.}
  {\bf C47} (2006) 337--346.

\bibitem{Dohmen:1993mp}
{\bf SINDRUM II Collaboration.} Collaboration, C.~Dohmen et~al., {\it {Test of
  lepton flavor conservation in $mu \to e$ conversion on titanium}},  {\em
  Phys.Lett.} {\bf B317} (1993) 631--636.

\bibitem{PRIME}
{\bf The PRIME working group} Collaboration, ``Search for the $\mu \to e$
  conversion process at an ultimate sensitivity of the order of $10^{-18}$ with
  prism.'' unpublished; LOI to J-PARC 50-GeV PS, LOI-25,
  \url{http://www-ps.kek.jp/jhf-np/LOIlist/pdf/L25.pdf}.

\bibitem{Davier:2010nc}
M.~Davier, A.~Hoecker, B.~Malaescu, and Z.~Zhang, {\it {Reevaluation of the
  Hadronic Contributions to the Muon g-2 and to alpha(MZ)}},  {\em Eur.Phys.J.}
  {\bf C71} (2011) 1515, [\href{http://xxx.lanl.gov/abs/1010.4180}{{\tt
  arXiv:1010.4180}}].

\bibitem{Pospelov:2005pr}
M.~Pospelov and A.~Ritz, {\it {Electric dipole moments as probes of new
  physics}},  {\em Annals Phys.} {\bf 318} (2005) 119--169,
  [\href{http://xxx.lanl.gov/abs/hep-ph/0504231}{{\tt hep-ph/0504231}}].

\bibitem{Baron:2013eja}
{\bf ACME Collaboration} Collaboration, J.~Baron et~al., {\it {Order of
  Magnitude Smaller Limit on the Electric Dipole Moment of the Electron}},
  \href{http://xxx.lanl.gov/abs/1310.7534}{{\tt arXiv:1310.7534}}.

\bibitem{Beringer:1900zz}
{\bf Particle Data Group} Collaboration, J.~Beringer et~al., {\it {Review of
  Particle Physics (RPP)}},  {\em Phys.Rev.} {\bf D86} (2012) 010001.

\bibitem{ATLAS:2013hma}
{\bf ATLAS} Collaboration, {\it {Search for Type III Seesaw Model Heavy
  Fermions in Events with Four Charged Leptons using 5.8 fb$^{-1}$ of $\sqrt{s}
  =$ 8 TeV data with the ATLAS Detector}},  Tech. Rep. ATLAS-CONF-2013-019,
  ATLAS-COM-CONF-2013-009, CERN, Geneva, 2013.

\bibitem{Alwall:2011uj}
J.~Alwall, M.~Herquet, F.~Maltoni, O.~Mattelaer, and T.~Stelzer, {\it {MadGraph
  5 : Going Beyond}},  {\em JHEP} {\bf 1106} (2011) 128,
  [\href{http://xxx.lanl.gov/abs/1106.0522}{{\tt arXiv:1106.0522}}].

\bibitem{Sjostrand:2006za}
T.~Sjostrand, S.~Mrenna, and P.~Z. Skands, {\it {PYTHIA 6.4 Physics and
  Manual}},  {\em JHEP} {\bf 0605} (2006) 026,
  [\href{http://xxx.lanl.gov/abs/hep-ph/0603175}{{\tt hep-ph/0603175}}].

\bibitem{deFavereau:2013fsa}
J.~de~Favereau, C.~Delaere, P.~Demin, A.~Giammanco, V.~Lemaître, et~al., {\it
  {DELPHES 3, A modular framework for fast simulation of a generic collider
  experiment}},  \href{http://xxx.lanl.gov/abs/1307.6346}{{\tt
  arXiv:1307.6346}}.

\bibitem{Carena:2012xa}
M.~Carena, I.~Low, and C.~E. Wagner, {\it {Implications of a Modified Higgs to
  Diphoton Decay Width}},  {\em JHEP} {\bf 1208} (2012) 060,
  [\href{http://xxx.lanl.gov/abs/1206.1082}{{\tt arXiv:1206.1082}}].

\bibitem{ALEPH:2005ab}
{\bf ALEPH Collaboration, DELPHI Collaboration, L3 Collaboration, OPAL
  Collaboration, SLD Collaboration, LEP Electroweak Working Group, SLD
  Electroweak Group, SLD Heavy Flavour Group} Collaboration, S.~Schael et~al.,
  {\it {Precision electroweak measurements on the $Z$ resonance}},  {\em
  Phys.Rept.} {\bf 427} (2006) 257--454,
  [\href{http://xxx.lanl.gov/abs/hep-ex/0509008}{{\tt hep-ex/0509008}}].

\bibitem{BenekeTalk}
M.~Beneke, ``{Flavour physics }.''
\newblock {Talk given at the Latsis Symposium in Zurich, 6 June 2013}.

\end{thebibliography}\endgroup

\end{document}